# New read-out electronics for ICARUS-T600 liquid Argon TPC. Description, simulation and tests of the new front-end and ADC system

**The ICARUS/NP01 Collaboration**

L. Bagby[a], B. Baibussinov[b], V. Bellini[c], M. Bonesini[d], A. Braggiotti[b,e], L. Castellani[b], S. Centro[b], T. Cervi[f], A.G. Cocco[g], F. Fabris[b], A. Falcone[f,1], C. Farnese[b,2], A. Fava[a,b], F. Fichera[c], D. Franciotti[h], G. Galet[b], D. Gibin[b], A. Guglielmi[b], R. Guida[b], W. Ketchum[a], S. Marchini[b], A. Menegolli[f], G. Meng[b], G. Menon[b], C. Montanari[f,i], M. Nessi[i], M. Nicoletto[b], R. Pedrotta[b], P. Picchi[f], F. Pietropaolo[b,i], G. Rampazzo[b], A. Rappoldi[f], G.L. Raselli[f], M. Rossella[f], C. Rubbia[h,i,j,3], A. Scaramelli[f], F. Sergiampietri[k], M. Spanu[f], M. Torti[f], F. Tortorici[c], F. Varanini[b], S. Ventura[b], C. Vignoli[h], A. Zani[i], P.G. Zatti[b]

[a] *Fermi National Accelerator Laboratory, Batavia, Illinois, USA*
[b] *Dipartimento di Fisica e Astronomia "G. Galilei", Università di Padova and INFN, Padova, Italy*
[c] *Dipartimento di Fisica e Astronomia, Università di Catania and INFN, Catania, Italy*
[d] *Dipartimento di Fisica "G. Occhialini", Università di Milano-Bicocca and INFN Milano-Bicocca, Milano, Italy*
[e] *Istituto di Neuroscienze, CNR, Padova, Italy*
[f] *Dipartimento di Fisica, Università di Pavia and INFN, Pavia, Italy*
[g] *Dipartimento di Scienze Fisiche, Università Federico II di Napoli and INFN, Napoli, Italy*
[h] *INFN - Laboratori Nazionali del Gran Sasso, Assergi, Italy*
[i] *CERN, Geneva, Switzerland*
[j] *GSSI, L'Aquila, Italy*
[k] *INFN, Pisa, Italy*

*E-mail*: christian.farnese@pd.infn.it

---

[1] Now at University of Texas, Arlington, Texas, USA
[2] Corresponding author
[3] Spokesperson

ABSTRACT: The ICARUS T600, a liquid argon time projection chamber (LAr-TPC) detector mainly devoted to neutrino physics, underwent a major overhauling at CERN in 2016-2017, which included also a new design of the read-out electronics, in view of its operation in Fermilab on the Short Baseline Neutrino (SBN) beam from 2019. The new more compact electronics showed capability of handling more efficiently the signals also in the intermediate Induction 2 wire plane with a significant increase of signal to noise (S/N), allowing for charge measurement also in this view. The new front-end and the analog to digital conversion (ADC) system are presented together with the results of the tests on 50 liters liquid argon TPC performed at CERN with cosmic rays.

**Introduction**

The LAr-TPC, an innovative detection technique that allows to accurately identify and reconstruct each ionizing track, has been proposed in 1977 [1] as alternative to Cerenkov radiation detectors. With the continue effort of ICARUS Collaboration and INFN support, the LAr-TPC technique has been taken to the full maturity with the large LAr mass T600 detector successfully operated in 2010-2013 at the LNGS underground laboratories exposed to CNGS beam and cosmic rays with a live-time > 93% [2].

ICARUS T600 consists of two identical modules filled with 760 t of ultra-pure liquid argon each one housing on both long sides TPC chambers separated by a central common cathode with 1.5 m drift length. A 500 V/cm uniform electric field allows for drifting without distortions the ionization electrons produced by charged particles along their path to three parallel read-out wire planes, oriented at $0^0, \pm 60^0$ with respect the horizontal direction facing the drift volume, ~54000 wires in total, 3 mm pitch and plane spacing. Induced signals in the first two Induction wire planes and the electron charge signals on the last Collection planes allow for measuring three independent event projections which are combined in a full 3D reconstruction of any ionizing event with a ~1mm space resolution. Moreover, the charge signal detected in the Collection view, proportional to the deposited energy, allows for the calorimetric measurement of the particle energy. To prevent the absorption of the drifting electrons by electronegative elements, which would result in a reduction of the signal, the residual LAr impurities are kept at the exceptionally low level <50 $O_2$ ppt (corresponding to an electron lifetime of 7 ms) by continuously filtering both liquid and gas argon [3]. A photomultiplier system installed behind the wire planes detects the scintillation light emitted by charged particles [4]. The drift time of each ionization signal combined with the 1.6 mm/µs electron drift velocity provides the position of the track along the drift coordinate. A detailed description of the ICARUS detector can be found elsewhere [5].

The three years long run of Icarus at LNGS demonstrated its remarkable event identification and measurement capabilities investigating the presence of the LSND-like $v_\mu$–$v_e$ oscillations in the CNGS neutrino beam [6]. Moreover a sample of atmospheric neutrino interactions in the 0.2–2 GeV energy range was collected to qualify the detector performance in the energy range of interest for the forthcoming operation of ICARUS-T600 in the SBN program at Fermilab aiming to definitely clarify the LSND effect [7].

The ICARUS front-end amplifier and ADC system, used in the LNGS run, performed efficiently with an extremely good S/N ratio that allowed for collecting some thousands neutrino and cosmic events with unprecedented quality. However, the signal shaping chosen at the time presented some limitations on signals produced by the intermediate Induction 2 wire plane in case of dense showers. The overhauling of ICARUS T600 performed at CERN in preparation for its operation at Fermilab, gave the opportunity of designing new electronics, which integrates the DAQ boards onto the signal flanges. Moreover, overhauling gave the opportunity of designing a new more compact front-end amplifier with a different signal shaping. This new design showed to be capable of handling more efficiently also the Induction 2 signals, with a significant increase of S/N allowing for measurement of deposited energy also in this view. In this paper, the new front-end and the AD conversion system will be presented together with results of the tests performed at CERN with the 50 liters liquid argon LAr-TPC [7].

## 1 Amplifier architecture

The adopted architecture is the same of the original design [5]: a Radeka-like amplifier with 3 parallel jFet as input stage, followed by an unfolded cascode integrated in a dual channel BiCMOS ASIC. Multiple jFet (BF861C) at input, were adopted to achieve a total $g_m$ in the order of 45 mS. The ASIC custom design was chosen for performance uniformity and compactness. To improve this latter a new even smaller package was chosen. The input stage is followed by a shaper (a classic zero-pole cancellation filter) and baseline restorer with a peaking time of 0.6 $\mu s$ in response of a delta-like input current. The basic scheme is given in Figure 1.

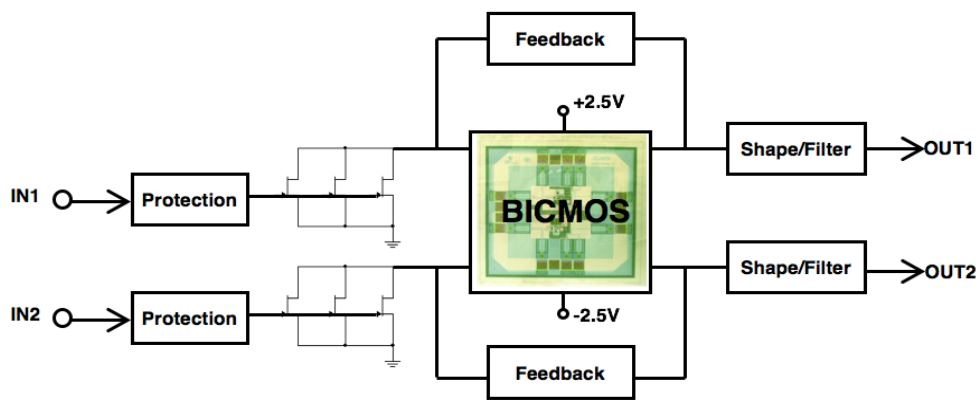

**Figure 1** Block diagram of the front-end amplifier.

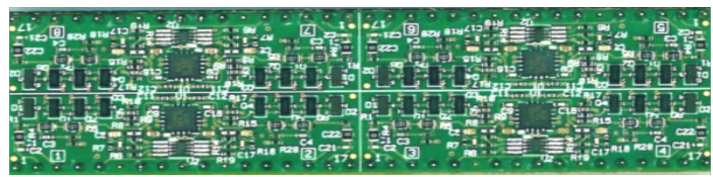

**Figure 2** PCB housing 8 amplifiers with simmetrical layout.

The same shaping time was chosen both for Induction and Collection signals. In this way, the bipolar shape of the Induction signals is preserved, allowing for proper treatment and reconstruction also for dense showers, as it will be shown on the real data collected with the test TPC at CERN.

Eight amplifiers are mounted on a single board (81 x 20 mm$^2$) shown in Figure 2. A symmetric layout for each couple of the eight channels was adopted to keep uniform the effects due to layout stray capacitances. Eight such small amplifier boards fit in each of the 8 connectors on the new A2795 CAEN motherboard, shown in Figure 3, for a total of 64 channels. The A2795 board was designed, engineered, and built by CAEN, in collaboration with ICARUS team and according to the experiment requirements. A throughput exceeding 10 Hz was realized by a modern switched I/O system where transactions are carried out over optical Gigabit/s serial links. Nine motherboards are housed in a special crate mounted onto the feedthrough flange (INFN proprietary design) designed for the transmission of the TPC wire signals (see Figure 8). This was an important achievement reducing the volume of the front-end electronics for each flange, serving 576 channels, from a volume of ~600 liters (a full rack) to 10 liters.

After the zero-pole cancellation filter, in front of the serial ADC, a Bessel filter due to its linear phase response preserves the area of the filtered signal in the passband. Such a filter interfaces each amplifier with its serial 12 bit (Least Significant Bit LSB = 0.8 mV, 400 ns sampling time) ADC (AD7276BUJZ) mounted onto the A2795 board. A dedicated ADC for each channel avoids any multiplexing also implementing synchronous conversion. Data buffering, digital processing, and transmission onto optical link are committed to a programmable FPGA (Altera Cyclone 5 GX).

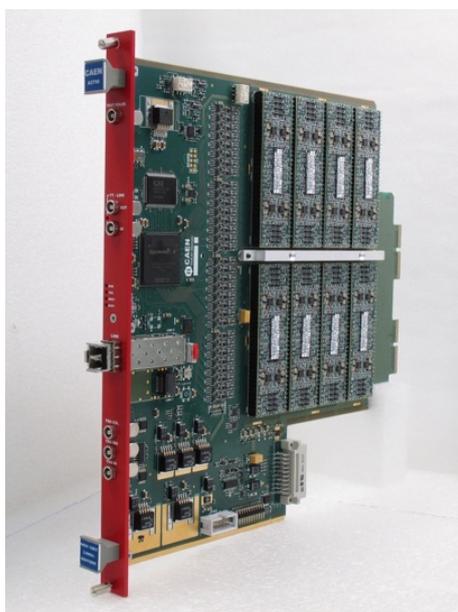

**Figure 3** A2795 custom board housing 64 amplifiers (far end), AD converter, digital control, and optical link (front panel).

## 2  Amplifier simulations

The simulation of the new electronic chain was performed with SPICE to model its response function, gain and noise performance.

The amplifier response with zero detector capacitance (30 pF amplifier capacitance) to an injected current 100 fC$_*$ $\delta(t)$ gives an output voltage with a peak time $\tau$ = 0.6 μs and amplitude of 1.64 V, where $\delta(t)$ is the derivative of the Heaviside function with dimension of [time$^{-1}$]. Increasing the value of the input capacitance from 30 to 500 pF would result in a corresponding reduction of the signal amplitude, as shown in Figure 4 at the input and output of the Bessel filter.

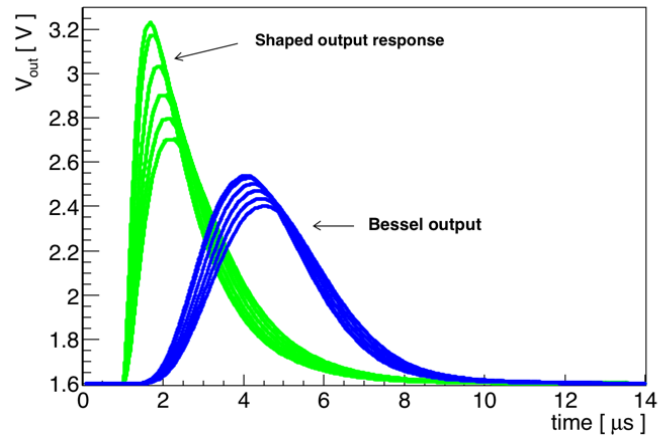

**Figure 4** Shaped output response (green) at 100fC∗δ(t) for Cd = 30, 100, 200, 300, 400, 500 pF, and corresponding Bessel output (blue).

Unlike the ideal case of infinite open loop amplification ($A_{OL}$), this is due to the non-zero input impedance of the front end, $Z_{inf} \approx 1$ kΩ. However this is not relevant because the full input current will be eventually integrated as series of appropriate impulsive currents $q\,\delta(t)$ and the charge information will be obtained from the signal area.

The frequency response of the open loop and feedback gains ($A_{OL}$, $A_F$) of the front-end, before zero-pole cancellation, are $A_{OL}$ = 140 db and $A_F$ = 75 db with a 20 db/dec slope in the 50 kHz - 500 MHz range as shown in Figure 5.

The time domain response after zero-pole cancellation, to an injected current $q\,\delta(t)$ is

$$V_{OUT}(t) = (R/\tau)\ q\ t/\tau\ e^{-t/\tau} = (q/C)\ x\ e^{-x} \qquad (2.1)$$

where R and C are depending on the circuit, $\tau = 0.6$ μs and $x = t/\tau$.

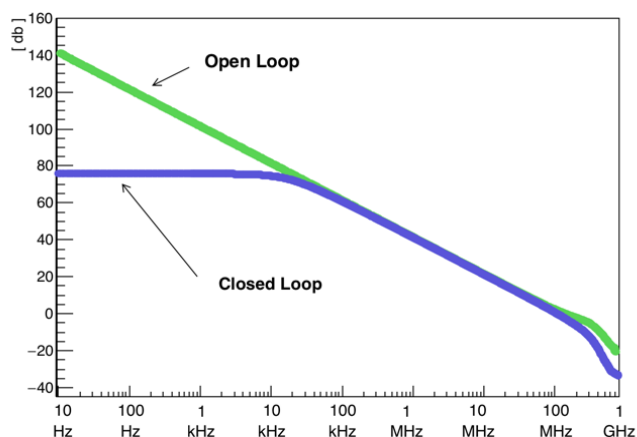

**Figure 5** Frequency characteristics: open loop (green) and closed loop (blue).

The signal $V_{OUT}(t)$ exhibits at $t = \tau$ the maximum value:

$$V_{OUTMAX}(\tau) = R/\tau\ q/e = q/C\ 1/e. \qquad (2.2)$$

For any kind of input current, the signal at the output of the preamplifier-shaper is the convolution of the input signal with the response function of the amplification chain:

$$V_{out}(t) = \frac{1}{C}\int_{-\infty}^{t} i(z) \frac{(t-z)}{\tau} e^{\frac{-(t-z)}{\tau}} dz . \qquad (2.3)$$

In case of an impulsive current (e.g. that of a test pulse) $i_{tp} = q_{tp}\delta(t)$, the output signal becomes

$$V_{out,tp}(t) = \frac{q_{tp}}{C} \frac{t}{\tau} e^{-\frac{t}{\tau}} \qquad (2.4)$$

and its area is proportional to the injected charge ($\int_{-\infty}^{\infty} x\, e^{-x} dx = 1$):

$$Area_{tp} = \sum_i V_{out,tp}(i)\, \Delta t = \frac{q_{tp}}{C}\int_{-\infty}^{\infty} \frac{t}{\tau} e^{\frac{-t}{\tau}} dt = \frac{q_{tp}}{C} \tau. \qquad (2.5)$$

Since any real input current signal recorded from the detector can be constructed as a sequence of impulsive signals with the appropriate intensity, the associated total charge can be extracted, independently from the signal shape, from the integrated area of $V_{out}$ as:

$$Area_{sig} = \sum_i V_{out,sig}(i)\, \Delta t = \frac{1}{C}\int_{-\infty}^{\infty} dt \int_{-\infty}^{t} i(z)\frac{(t-z)}{\tau} e^{\frac{-(t-z)}{\tau}} dz = \frac{Q_{sig}}{C}\tau, \qquad (2.6)$$

hence as a general rule, applying test pulse calibration, the resulting charge value is:

$$Q_{sig} = Area_{sig}\, \frac{q_{tp}}{Area_{tp}}. \qquad (2.7)$$

The formula (2.7), which has been calculated after the zero-pole cancellation filter, is still valid at the Bessel output because it concerns only areas ratio and the Bessel filter has the peculiarity of preserving the shaped area.

The $V_{out}$ signal is digitized by the ADC with LSB = 0.8 mV amplitude at 400 ns sampling time. For 100 fC input charge, at 30 pF input capacitance, the area of the output signal as calculated in the simulation is $3.07\, 10^{-6}$ Vs. Being an ADC least count area $0.8 \cdot 400\, 10^{-12}$ Vs, the 100 fC correspond to 9566 ADC elementary areas. Therefore the charge corresponding to 1 ADC count is:

$$q_{LSB} = 0.0104 \text{ fC} \quad \text{or} \quad 65 \text{ electrons} \qquad (2.8)$$

While the $Area_{sig}$ is 9566 ADC counts the pulse-height at the peak time is $Q_{peak}$ = 1181 counts. The corresponding ratio for a δ-like current, is then 9566/1181 = 8.1. In other words, 0.8 mV peak signal is equivalent to 8.1 ADC area counts or ~ 526 electrons. However while the collected charge is still represented by the area of the signal, the pulse-height is reduced by increasing the input capacitance as shown in Figure 4, and the area-to-peak ratio shifts from 8.1 to 8.9 for an input capacitance of 30 to 500 pF respectively (Figure 6).

The RMS noise was simulated as the square root of the integral of the square power density [$V^2$/Hz] calculated over the bandwidth of the amplifier taking into account the cut-off due to the sampling time. The resulting Equivalent Noise Charge (ENC) as a function of the total input capacitance is shown in Figure 7: the predicted noise increases with a 2.3 electrons/ pF slope.

The Bessel filter output is ±1.26 V, with a maximum input charge of ±160 fC that does not saturate the input stage for an injection time of 2 μs typical of the T600. According to the 3.3 V ADC full-scale, a total charge of ±160 fC can be managed using 76% of the 12 bit range.

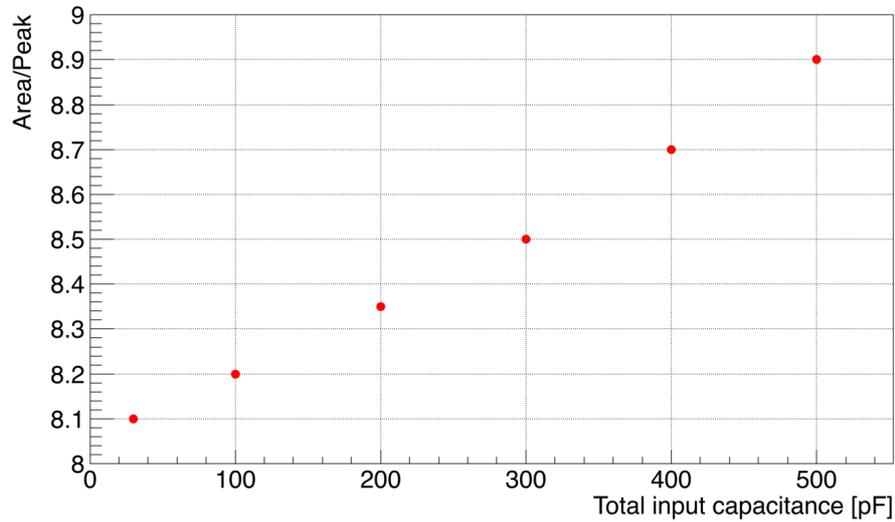

**Figure 6** The simulated Area/Peak signal ratio as a function of total input capacitance.

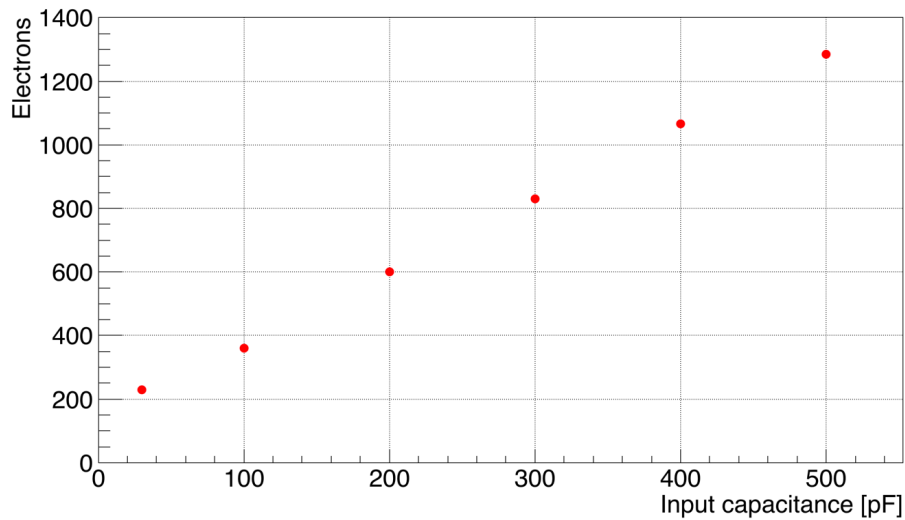

**Figure 7** Simulated output ENC noise versus input capacitance.

## 3  Amplifier tests

Tests were performed at CERN, with the 50 liters ICARUS LAr-TPC test facility [7], characterized by a drift field of 500 V/cm on a drift length of 52 cm. The chamber is approximately 32 x 32 cm$^2$ wide and has three wire planes with 4 mm separation, 128 wires per plane, and a wire pitch p = 2.54 mm. The first plane (grid) facing the drift volume is not read out; the wires of the intermediate

Induction are parallel to those of the grid, while those of the last Collection plane are orthogonal. The grid and the Collection planes were biased up to ±400 V (negative for grid, positive for Collection) while keeping the Induction at 0 V. The Induction and Collection wires were connected to four A2795 boards inserted in a specially designed mini-crate installed on the feedthrough flange through 2.7 m long twisted pair cables, 57 pF/m capacitance, of the type used in ICARUS-T600 (Figure 8). On each board the first group of 32 channels was used for the Induction signals and the other 32 for the Collection, both read out through two optical fibers.

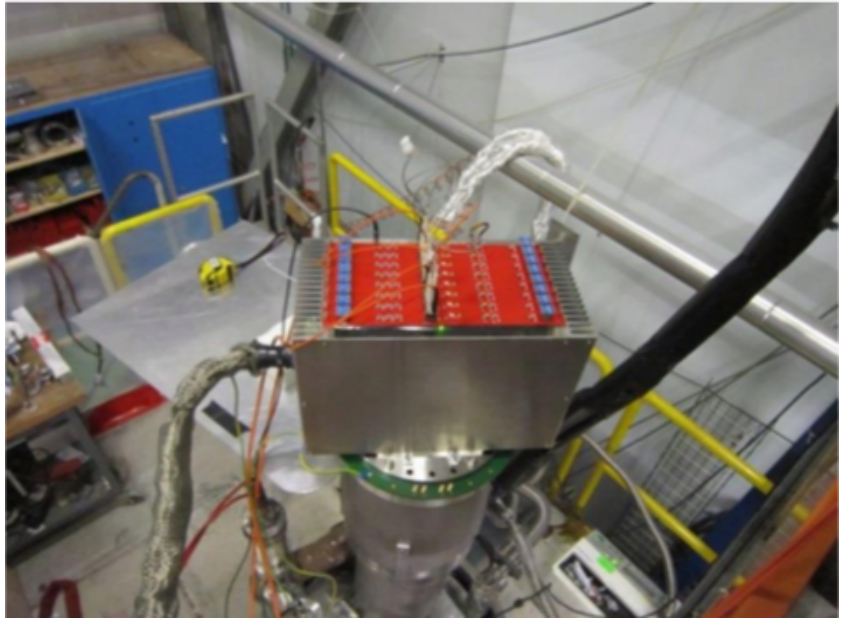

**Figure 8** Test set up on the 50 liters TPC at CERN. Four of nine boards are connected to 128 Collection and Induction wires.

All the channels are equipped with a 1.75 pF (±1%) very precise test capacitance obtained with a copper strip in an inner layer of the PCB frame of the TPC wires, orthogonal to the contact tracks joining the wires to the cable connectors.

Gain and linearity of the full electronic chain, which includes the preamplifier, the shaper, the Bessel filter and the 12 bit ADC converter, were evaluated by injecting on the test capacitance fast charge pulses (0.1 ns rise time) in the 17.5 —105 fC range. The resulting pulse area distribution (4732 ± 115 ADC counts) is shown as a function of the channel number in Figure 9 for an injected charge of 52.5 fC (327600 electrons). The gain turns out $q_{LSB}$ = 69.2±1.7 electrons /ADC count to be compared with the 65 electron/ADC count as predicted by the simulation, the 2.5% uncertainty reflecting the small non-uniformity of the electronic chain. The dependence of the signal area on the rise time was verified to be < 2% by injecting test pulses with rise time in the 0.1 – 4 μs range (Figure 10). In conclusion the total charge of any signal $Q_{tot}$ is then given by

$$Q_{tot} = \Sigma_i \, q_{LSB} \, ADC_i \qquad (3.1)$$

where $ADC_i$ is the ADC count at the i-sample.

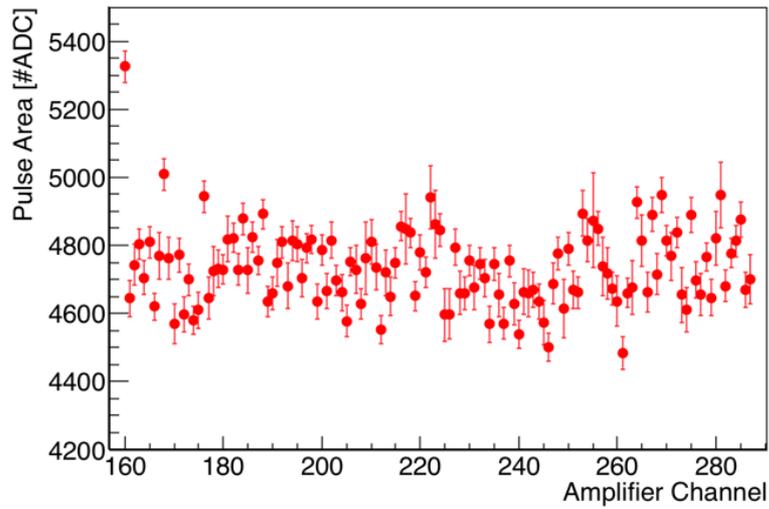

**Figure 9** Pulse area as a function of the amplifier channel in Collection view (52.5 fC injected charge, 0.1 μs risetime).

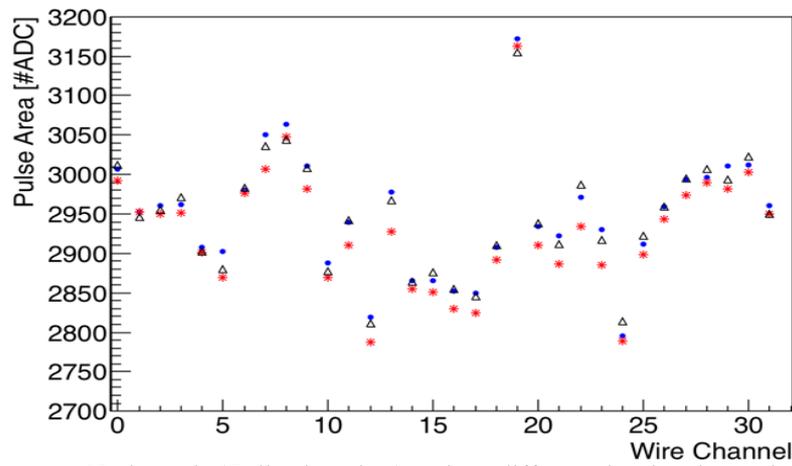

**Figure 10** Pulse areas on 32 channels (Collection view) at three different rise time input signals (0.1, 2.0 and 4.0 μs in red, blue, black respectively) for 32 fC injected charge.

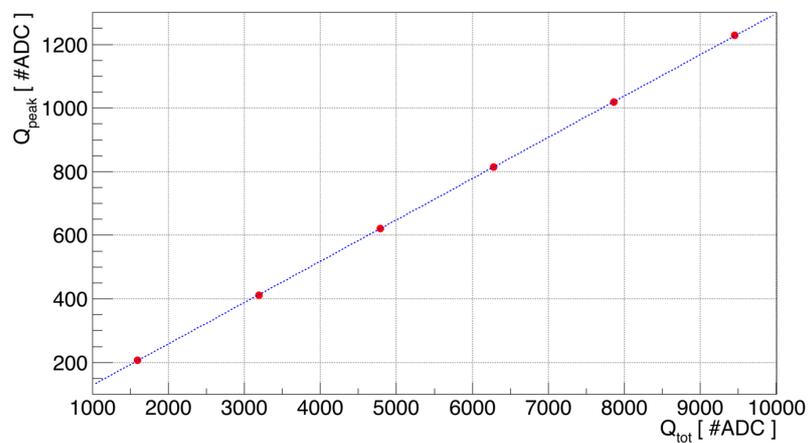

**Figure 11** $Q_{peak}$ versus different δ-like input charges $Q_{tot}$.

Finally, the injection of δ-like charge pulses in the 17.5 fC—105 fC range allowed to verify the gain linearity within 0.5%, and to test the linear relation between the total charge $Q_{tot}$ (signal area) and the signal pulse-height $Q_{peak}$ (see Figure 11), resulting in

$Q_{tot}$= 7.7 $Q_{peak}$.                                                                                                          (3.2)

The measured 7.70 ± 0.06 ratio is consistent with the ~8.2 simulated value for the ~130 pF total input capacitance, including the cables, as quoted in Figure 6.

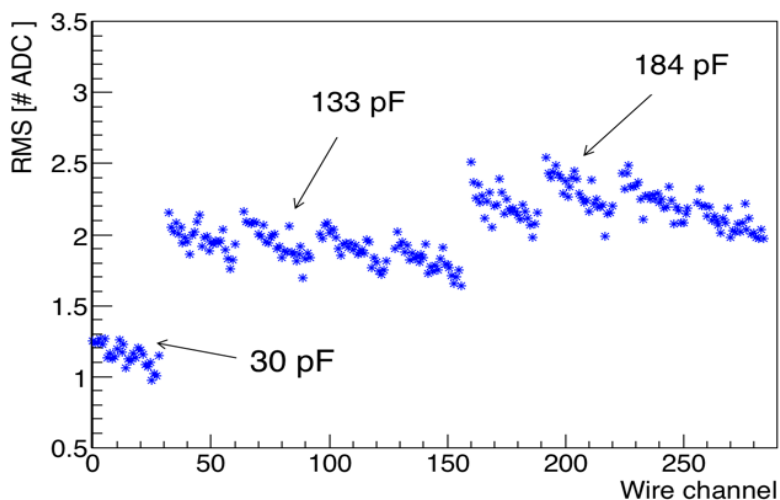

**Figure 12** RMS noise in ADC counts for the three input capacitance configurations as described in the text.

Noise measurements were performed in three set-up conditions, characterized by different input capacitances (Figure 12): 30 pF (JFet only), 133 pF and 184 pF as introduced by the addition of 1.8 m and 2.7 m cable length respectively. A clear channel-to-channel noise pattern regularly repeating in each board, related to the A2795 layout, is recognized. The smallest noise contribution is in the last channel in a set of 32. Moreover, a coherent noise on each board, visible as a texture in the event images (see for example Figure 14 and Figure 15), was evaluated from the average, for each time sample, of the 32 wire waveforms in the same board **(**Table 1**)**.

**Table 1** RMS noise measurements (pulse-height) for the three different input capacitances quoted in the text. Column 3: average noise over 64 channels in a board; Column 4: the minimum noise averaged over the boards with the same Cd; Column 5: average coherent noise; Column 6 and 7: intrinsic channel noise, to be compared with simulation. The conversion factors in column 2 are determined from Figure 6 taking into account relation (2.8).

| Cd (pF) | Conversion factor | Electronic noise | | | | |
|---|---|---|---|---|---|---|
| | Electrons/ ADC peak count | Average (counts) | Minimum (counts) | Coherent (counts) | Intrinsic (counts) | Intrinsic (electrons) |
| 30 | 526 | 1.15±0.08 | 1.08±0.04 | 0.32±0.02 | 1.03±0.04 | 542±21 |
| 133 | 536 | 1.90±0.11 | 1.82±0.06 | 0.81±0.05 | 1.63±0.08 | 874±43 |
| 184 | 542 | 2.21±0.13 | 2.13±0.08 | 0.98±0.07 | 1.89±0.11 | 1024±60 |

For sake of comparison with the single channel simulation, the contribution to the noise introduced by the A2795 layout was removed by selecting the least noisy channel in each board (column 4 of Table 1). After subtracting in quadrature the coherent contribution from the value associated to this channel (Column 5 of Table 1) the intrinsic noise of a single channel is obtained (column 6, 7 of Table 1). Both the average and intrinsic measured noise are shown in Figure 13 as a function of the total input capacitance. A noise slope of 3.89 ± 0.48 electrons/pF in the first case, reduced to 3.17 ± 0.33 electrons/pF for the intrinsic noise, is closer to 2.3 electrons/pF value obtained from the simulation.

While the board layout effects and coherent noise cannot be included in the simulation, the results in Table 1 provide a reasonable picture of the front-end electronics from the point of noise slope. The different offset of the noise in Figure 7 and Figure 13 indicates that some parallel noise sources are not properly taken into account in the simulation.

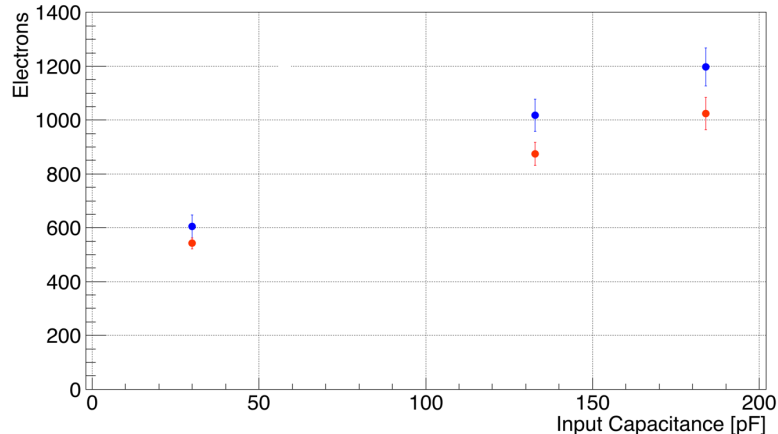

**Figure 13** Noise slope for the three quoted input capacitances: blue and red values refer to average and intrinsic noise respectively as quoted in Table 1.

According to the 3 mm wire pitch and plane spacing of ICARUS-T600, a signal to noise ratio S/N ≈15000/1200 ~12 can be roughly expected for ~15000 electrons signal released on a single TPC wire by a m.i.p. particle in 3 mm LAr [5].

## 4   Tests with cosmic rays

Cosmic ray events were collected at CERN with the 50 liters ICARUS LAr-TPC equipped with the new electronics, at the nominal drift field of 500 V/cm. An external trigger was set up based on the coincidence of two external plastic scintillator slabs and an internal Hamamatsu 8" photomultiplier coated with TPB wavelength shifter [5]. It selected mainly cosmic muon tracks inclined at about $45^0$ with respect to the drift direction, spanning most of the wires in both the Induction and Collection planes (an example is shown in Figure 14). Moreover a fraction of the triggered events consisted of electromagnetic showers (Figure 15) induced by cosmic rays.

Muon tracks were used to calibrate the electronic chain with minimum ionizing particles. Measurements were performed to reach an almost full transparency to the drifting electrons for both the screen grid and the Induction plane by varying the biasing voltage in the ±275, ±400 V range on the screen grid and the Collection plane while keeping the induction plane at 0 V. Landau dE/dx distributions of the charge deposited in LAr by minimum ionizing particles (dE/dx ~ 2.1 MeV/cm in LAr) were obtained ( Figure 16 ) by selecting through-going muon tracks reconstructed in 3D and calculating the signal area on the wires of the Collection plane with the previously described procedure. Small corrections, due to finite electron lifetime, were also introduced as measured from the same muon track sample.

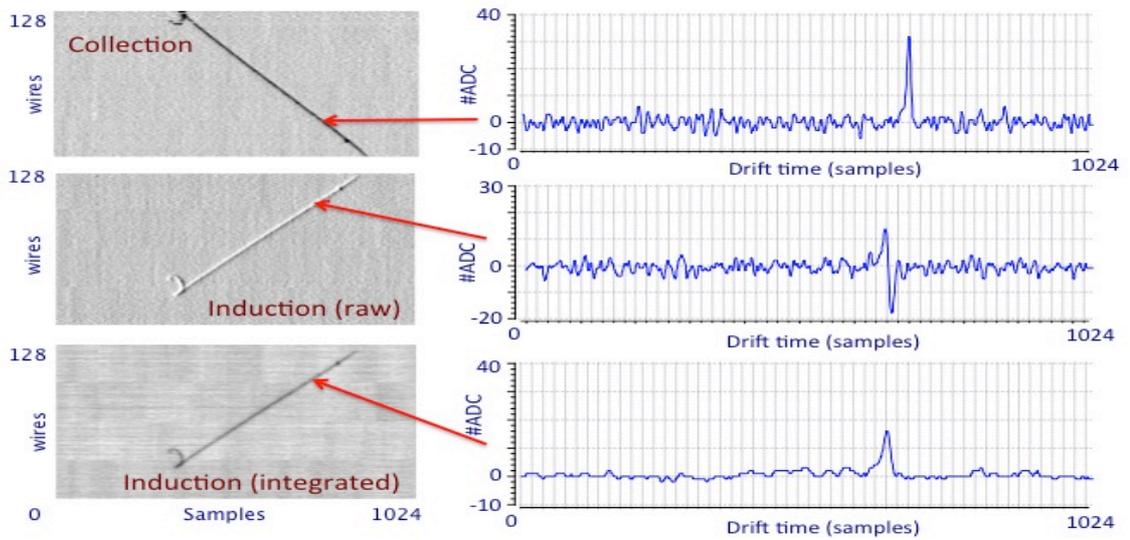

**Figure 14** Example of single muon tracks at ~$45^0$ with respect to the drift direction collected with the 50 liters ICARUS LAr-TPC and the new electronics with ±400 V bias in Collection and grid wires. Induction signal is shown as raw data as well as after integration and filtering to recover unipolar features. The corresponding waveforms recorded on one channel are also reported on the right.

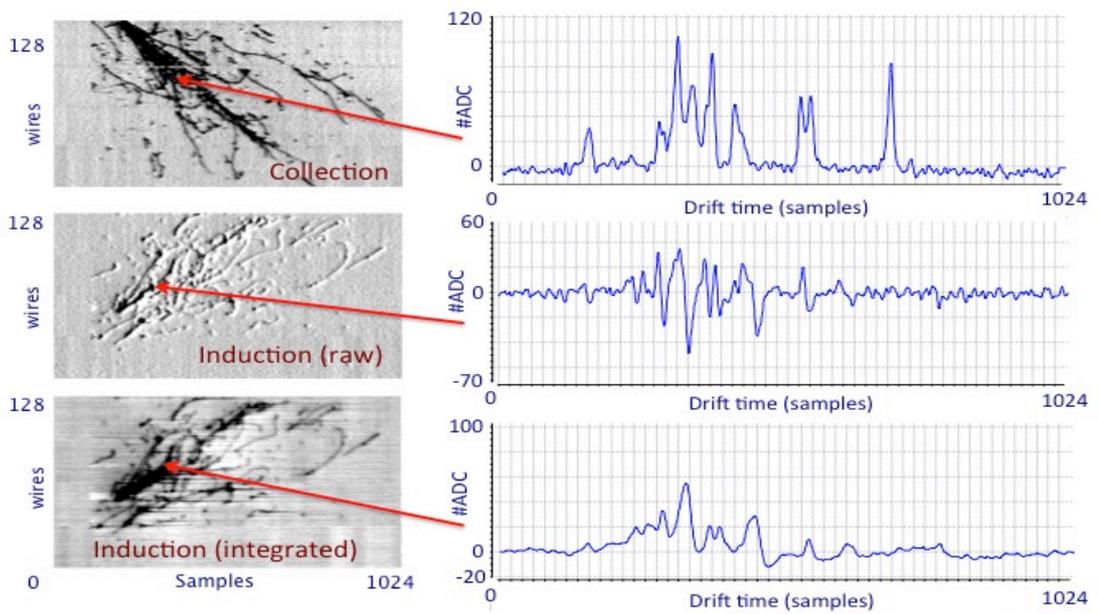

**Figure 15** Example of multi-prong event (electromagnetic shower) collected with the 50 liters ICARUS LAr-TPC and the new electronics with ±400 V bias in Collection and grid wires. Induction signal is shown as raw data as well as after signal integration to recover unipolar features. A corresponding waveform signal recorded on one channel is reported on the right.

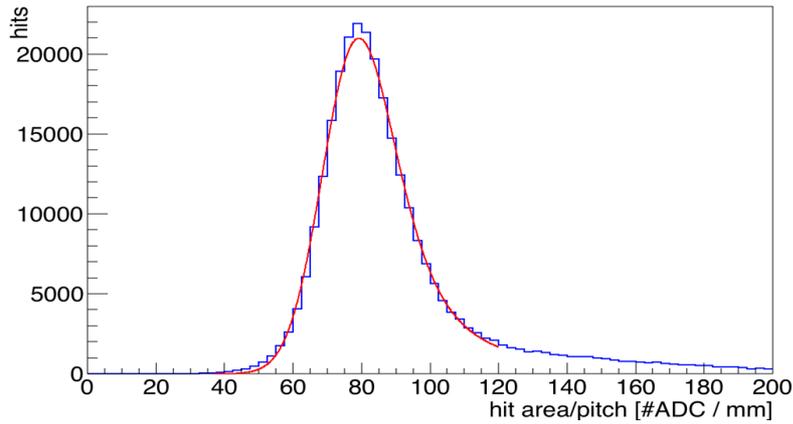

**Figure 16** Example of Landau dE/dx distribution from through-going muon tracks as measured on the Collection plane as fitted with a Landau-Gaussian convolution (±400 V/cm biasing voltage). The most probable value is 76 #ADC/mm. The width of the distribution, evaluated from the RMS of the fitting Gaussian and the scale parameter of the fitting Landau added in quadrature, result 12% of the most probable value, in agreement with the Landau and the electronic noise fluctuations.

The transparency of both the screen grid and Induction plane was determined as a function of the biasing voltage by studying the most probable value of the muon dE/dx Landau distribution, scarcely affected by track inclination and dE/dx relativistic rise. An almost full grid transparency is reached at 400 V for which the collected charge is ~76 (ADC counts)/mm (Figure 17).

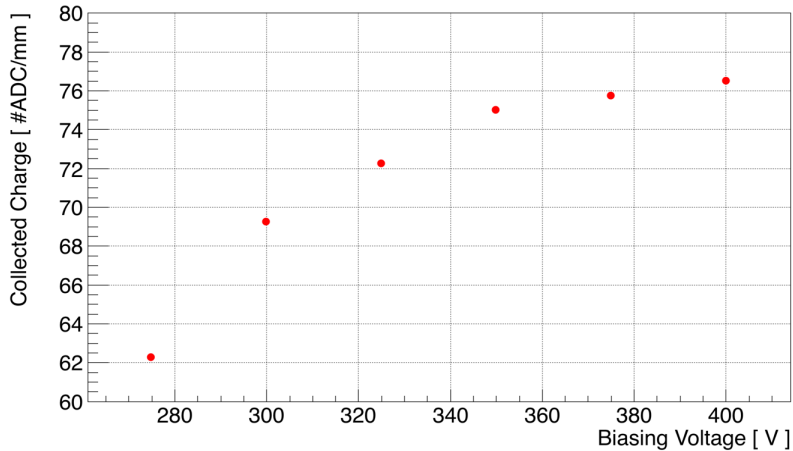

**Figure 17** The most probable value of the dE/dx distributions from through-going muon tracks as measured on the collection plane as a function of the wire planes biasing voltage.

At the nominal drift field of 500 V/cm the electron-ion recombination parameter is ~0.7 as measured by ICARUS at LNGS [4]. The dE/dx most probable value of 1.75 MeV/cm corresponds to 0.85 fC/mm. As a consequence, the electronic calibration of signal on Collection was measured to be:

C= 0.85 fC/mm/76 (ADC counts/mm) = 0.011 fC/ADC counts (~ 69 electrons/ADC)            (4.1)

in agreement with both test pulse data (68.4±1.6 electrons/ADC count) and simulations (65 electrons/ADC count).

While the Collection signal is simply proportional to the current i(t) collected on the wire at a given time, the bipolar nature of the Induction signal requires a specific treatment in order to extract a signal proportional to the charge. The Induction signal origins from both the contribution of the electron cloud approaching the Induction plane at a time t, and the one drifting away from the Induction wires, i.e. the incoming electrons at a previous time t-Δt, where Δt is the transit time between the two wire planes. An unipolar signal similar to the Collection one is obtained by integrating this differential signal over time and normalizing it for the average transit time Δt; the charge sensed by the Induction wires can be then recovered by a second integration. This two steps procedure, however, would enhance the low frequency noise components, requiring an additional filtering for the charge measurement.

As a first possible implementation of this scheme, a running sum integration of the Induction signal was performed followed by a baseline restoring based on the difference between the integrated signal itself and its average over ~30 μs, a much longer time scale compared to the few μs signal duration. The normalization factor Δt ~ 4 μs has been obtained from the average duration of Induction signals, after correcting for the track inclination. The outcome of this process is shown in the bottom images of Figure 14, Figure 15.

The adoption of the new optimized preamplifier in the front-end for both Induction and Collection wires with a fast signal shaping time to match the electron transit time in the wire spacing, prevents the signal undershoot as well as reducing the low frequency noise. The obtained unprecedented event image sharpness with an improved hit separation allows to resolve crowded and complex events both in Collection and Induction views. The recorded muon track (Figure 14) shows an average peak signal to RMS noise ratio S/N ≈30/2.2 ~14 and S/N ≈15/1.5 ~10 in Collection and Induction views after integration and filtering. The obtained values would further increase for ICARUS-T600 if the 3 mm different wire pitch and wire plane separation are accounted for. By a term of comparison, even if the bipolar treatment of the Induction signals prevents a completely homogeneous comparison, the corresponding S/N ratios as measured in T600 with the previous front-end electronics for similarly inclined muons were ~14 in Collection and ~7 in Induction.

As an initial evaluation of the potentialities of the dE/dx measurement in the Induction plane, the Landau distribution of the charge sensed by the Induction wires has been computed for the collected muons (Figure 18). The difference between Collection and Induction pulse areas is mainly due to the fact that only ~70% of drifting electrons charge is expected to be sensed on induction wires due to the geometry of the detector; a further small reduction of the signal height is due to the filtering procedure. The larger width of the distribution is attributed to the not fully filtered low frequency noise.

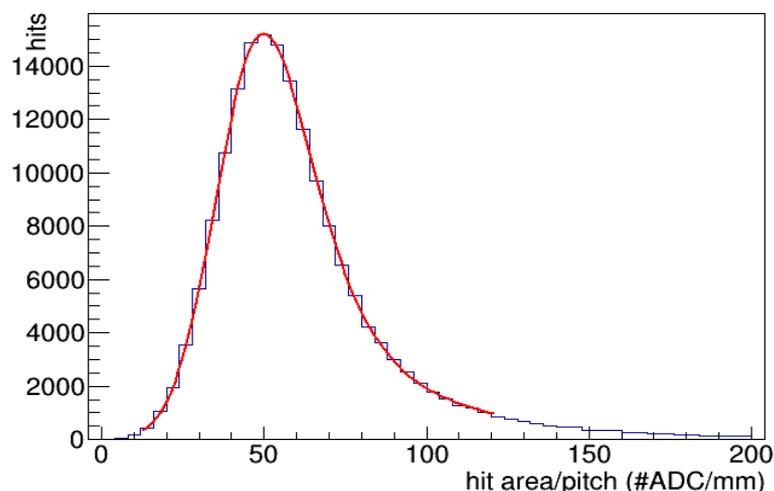

**Figure 18** Example of Landau dE/dx distribution from through-going muon tracks as measured on the Induction plane, fitted with a Landau-Gaussian convolution. Biasing voltage is ± 400 V/cm. The most probable value is 45 # ADC/mm. the percent width of the distribution as defined in Figure 15, is ~24%.

# 5 Conclusions

The new read-out electronics for ICARUS-T600 liquid argon TPC prepared in view of its operation in Fermilab on SBN beam includes a new design front-end, a serial 12 bit ADC system (one per channel) and serial bus architecture with optical links for Gigabit/s data transmission. In the new compact set-up both the analog and digital electronics are hosted directly on ad-hoc signal feed-through flange acting as electronic back-plane. The digital part is fully contained in a single high-performance FPGA in each board handling the ADC data. The same new preamplifier was adopted in the front-end for both Induction and Collection wires, with a faster shaping time to match the electron transit time in the wire plane spacing. Globally the throughput of the read-out system exceeds 10 Hz with optical Gigabit/s serial links.

The new ICARUS electronic chain was successfully tested with the 50 liters LAr-TPC exposed to cosmic rays. Beside its already well-established performance in the charge Collection wires a more efficient handling of signals in the intermediate Induction 2 wire plane with a significant increase of S/N was measured. The optimized preamp architecture resulted in unprecedented image sharpness of the events, with a better hit signal separation even in crowded and complex events like electromagnetic showers, both in Collection and Induction views.

The use of dedicated algorithms for the bipolar signals from Induction wires allows for measuring energy deposition in Induction view also, with a $\Delta E/E$ ~24% resolution on the single hit. At low energy Booster neutrino beam this feature will enable the recovering of leading lepton tracks travelling along the Collection wire direction for which the signal cannot be fully exploited for the dE/dx analysis and particle track separation, with an expected significant increase of the electron neutrino reconstruction efficiency [7].

## Acknowledgments

This work was funded by INFN in the framework of WA104/NP01 program finalized to the overhauling of ICARUS detector in view of its operation on SBN at Fermilab. The A2795 board was designed, engineered, and built by CAEN, in collaboration with ICARUS team and Electronics Service facilities of INFN, Padova. The strong contribution of A. Mati and A. Romboli from CAEN was essential.